\pdfoutput=1
\documentclass[prb,twocolumn,aps]{revtex4}
\usepackage{graphicx,graphics,color,epsfig}
\usepackage{bm}
\usepackage{amsmath}
\usepackage{amssymb}
\usepackage{epstopdf}

\makeatletter

\newcommand{\Rmnum}[1]{\expandafter\@slowromancap\romannumeral #1@}
\newcommand*{\rom}[1]{\expandafter\@slowromancap\romannumeral #1@}
\makeatother

\usepackage{color}

\begin{document}
\title{Robust $s_\pm$-wave pairing in a bilayer two-orbital model of pressurized La$_3$Ni$_2$O$_7$ without the $\gamma$ Fermi surface}

\author{Yi Gao}
\affiliation{Center for Quantum Transport and Thermal Energy Science, Jiangsu Key Lab on Opto-Electronic Technology, School of Physics and Technology, Nanjing Normal University, Nanjing 210023, China}

\begin{abstract}
We studied the superconducting pairing symmetry based on a newly constructed tight-binding model of La$_3$Ni$_2$O$_7$ under pressure, where the $\gamma$ band sinks below the Fermi level and does not form the Fermi surface. The superconducting pairing symmetry is $s_\pm$-wave and is robust against the variation of the interaction strength. In this model, although the $\gamma$ and $\delta$ bands are away from the Fermi level, the superconducting pairing function on them is not tiny. Instead, since the top of the $\gamma$ band and bottom of the $\delta$ band are both located at $\sim$500 meV away from the Fermi level, and they are almost nested by the peak structure in the spin fluctuation, thus by forming an anti-phase pairing function on them, these two bands act constructively to superconductivity. Finally with detailed derivation and numerical calculation, we demonstrate that the Fermi surface approximated Eliashberg equation may lead to deviation of the pairing symmetry.

\end{abstract}

\maketitle

\section{introduction}
In the year 2023, superconductivity with a transition temperature $T_c\approx80$ K is discovered in La$_3$Ni$_2$O$_7$ under pressure \cite{discovery}. In the high-pressure phase, it has an orthorhombic structure of $Fmmm$ space group, each unit cell contains two Ni atoms and the 3$d_{x^2-y^2}$ and 3$d_{z^2}$ orbitals of Ni dominate the energy bands close to the Fermi level. Soon after, various bilayer two-orbital tight-binding models were constructed to account for the electronic structure of this material under high pressure \cite{bilayermodel,s2,s5,s9,sd4,d2,d3,luhongyan,s11,sd3}. All of these models feature four bands, labeled as $\alpha$, $\beta$, $\gamma$ and $\delta$. The $\alpha$ and $\gamma$ bands are symmetric with respect to exchanging the two layers, while the $\beta$ and $\delta$ bands are anti-symmetric. $\alpha$, $\beta$ and $\gamma$ bands cross the Fermi level and form three Fermi surfaces while the $\delta$ band is above the Fermi level and unoccupied. The superconducting pairing symmetry has been studied based on these tight-binding models. It was found that, small differences in the details of the band structure may lead to a competition between the $s_\pm$- and $d$-wave pairing symmetries \cite{s1,s2,s3,s4,s5,s6,s7,s8,s9,s10,s11,s12,sd1,sd2,sd3,sd4,sd5,sd6,d1,d2,d3,luhongyan,sd7}. Recently, it was argued that, the above mentioned tight-binding models are fitted to the band structures calculated by the density functional theory (DFT), by using the generalized gradient approximation (GGA) for the exchange-correlation functional, which is, however, inaccurate \cite{jiangkun1,jiangkun2,jiangkun3}. Instead, Refs. \onlinecite{jiangkun1,jiangkun2,jiangkun3} adopted a more accurate hybrid exchange-correlation functional and constructed corresponding tight-binding model parameters. The main difference is, the $\gamma$ band shifts downward to 452 meV below the Fermi level, with only the $\alpha$ and $\beta$ bands forming two Fermi surfaces. Since the pairing symmetry relies on the detailed band structure in this material, therefore in this work, we investigate the superconducting pairing symmetry based on the newly constructed tight-binding model.

\section{method}

We adopt the bilayer two-orbital model of La$_3$Ni$_2$O$_7$ in the high-pressure phase proposed by Refs. \onlinecite{jiangkun1,jiangkun2,jiangkun3}. The two orbitals 3$d_{x^2-y^2}$ and 3$d_{z^2}$ of Ni are denoted as $x$ and $z$, respectively. The tight-binding part of the Hamiltonian can be written as $H_0=\sum_{\mathbf{k}\sigma}\psi_{\mathbf{k}\sigma}^{\dag}M_{\mathbf{k}}\psi_{\mathbf{k}\sigma}$, where $\psi_{\mathbf{k}\sigma}^{\dag}=(c_{\mathbf{k}1x\sigma}^{\dag},c_{\mathbf{k}2x\sigma}^{\dag},c_{\mathbf{k}1z\sigma}^{\dag},c_{\mathbf{k}2z\sigma}^{\dag})$ and
\begin{eqnarray}
\label{h0}
M_{\mathbf{k}}&=&\begin{pmatrix}
T_{\mathbf{k}}^{x}&t_{\bot}^{x}&V_{\mathbf{k}}&V_{\mathbf{k}}^{'}\\
t_{\bot}^{x}&T_{\mathbf{k}}^{x}&V_{\mathbf{k}}^{'}&V_{\mathbf{k}}\\
V_{\mathbf{k}}&V_{\mathbf{k}}^{'}&T_{\mathbf{k}}^{z}&t_{\bot}^{z}\\
V_{\mathbf{k}}^{'}&V_{\mathbf{k}}&t_{\bot}^{z}&T_{\mathbf{k}}^{z}
\end{pmatrix}.
\end{eqnarray}
Here $c_{\mathbf{k}lo\sigma}^{\dag}$ creates a spin $\sigma$ ($\sigma=\uparrow,\downarrow$) electron with momentum $\mathbf{k}$ in the layer $l$ ($l=1,2$) and orbital $o$ ($o=x,z$). In addition,
\begin{eqnarray}
T_{\mathbf{k}}^{x/z}&=&2t_{1}^{x/z}(\cos k_x+\cos k_y)+4t_{2}^{x/z}\cos k_x\cos k_y+\epsilon^{x/z},\nonumber\\
V_{\mathbf{k}}&=&2t_{3}^{xz}(\cos k_x-\cos k_y),\nonumber\\
V_{\mathbf{k}}^{'}&=&2t_{4}^{xz}(\cos k_x-\cos k_y).
\end{eqnarray}

By defining the symmetric and anti-symmetric basis with respect to exchanging the two layer indices as
\begin{eqnarray}
\label{symmetric_antisymmetric}
d_{\mathbf{k}So\sigma}&=&\frac{1}{\sqrt{2}}(c_{\mathbf{k}1o\sigma}+c_{\mathbf{k}2o\sigma}),\nonumber\\
d_{\mathbf{k}Ao\sigma}&=&\frac{1}{\sqrt{2}}(c_{\mathbf{k}1o\sigma}-c_{\mathbf{k}2o\sigma}),
\end{eqnarray}
the tight-binding Hamiltonian can be written as $H_0=\sum_{\mathbf{k}\sigma}\varphi_{\mathbf{k}\sigma}^{\dag}M^{'}_{\mathbf{k}}\varphi_{\mathbf{k}\sigma}$, where $\varphi_{\mathbf{k}\sigma}^{\dag}=(d_{\mathbf{k}Sx\sigma}^{\dag},d_{\mathbf{k}Sz\sigma}^{\dag},d_{\mathbf{k}Ax\sigma}^{\dag},d_{\mathbf{k}Az\sigma}^{\dag})$ and
\begin{eqnarray}
\label{M'k}
M^{'}_{\mathbf{k}}&=&\begin{pmatrix}
T_{\mathbf{k}}^{x}+t_{\bot}^{x}&V_{\mathbf{k}}+V_{\mathbf{k}}^{'}&0&0\\
V_{\mathbf{k}}+V_{\mathbf{k}}^{'}&T_{\mathbf{k}}^{z}+t_{\bot}^{z}&0&0\\
0&0&T_{\mathbf{k}}^{x}-t_{\bot}^{x}&V_{\mathbf{k}}-V_{\mathbf{k}}^{'}\\
0&0&V_{\mathbf{k}}-V_{\mathbf{k}}^{'}&T_{\mathbf{k}}^{z}-t_{\bot}^{z}
\end{pmatrix}.
\end{eqnarray}
$M^{'}_{\mathbf{k}}$ is block-diagonalized and its eigenvalues can be expressed as
\begin{eqnarray}
\label{eigenvalues}
E_{\mathbf{k}}^{\gamma}&=&\frac{1}{2}\{T_{\mathbf{k}}^{x}+t_{\bot}^{x}+T_{\mathbf{k}}^{z}+t_{\bot}^{z}\nonumber\\
&-&[(T_{\mathbf{k}}^{x}+t_{\bot}^{x}-T_{\mathbf{k}}^{z}-t_{\bot}^{z})^2+4(V_{\mathbf{k}}+V_{\mathbf{k}}^{'})^2]^{\frac{1}{2}}\},\nonumber\\
E_{\mathbf{k}}^{\alpha}&=&\frac{1}{2}\{T_{\mathbf{k}}^{x}+t_{\bot}^{x}+T_{\mathbf{k}}^{z}+t_{\bot}^{z}\nonumber\\
&+&[(T_{\mathbf{k}}^{x}+t_{\bot}^{x}-T_{\mathbf{k}}^{z}-t_{\bot}^{z})^2+4(V_{\mathbf{k}}+V_{\mathbf{k}}^{'})^2]^{\frac{1}{2}}\},\nonumber\\
E_{\mathbf{k}}^{\beta}&=&\frac{1}{2}\{T_{\mathbf{k}}^{x}-t_{\bot}^{x}+T_{\mathbf{k}}^{z}-t_{\bot}^{z}\nonumber\\
&-&[(T_{\mathbf{k}}^{x}-t_{\bot}^{x}-T_{\mathbf{k}}^{z}+t_{\bot}^{z})^2+4(V_{\mathbf{k}}-V_{\mathbf{k}}^{'})^2]^{\frac{1}{2}}\},\nonumber\\
E_{\mathbf{k}}^{\delta}&=&\frac{1}{2}\{T_{\mathbf{k}}^{x}-t_{\bot}^{x}+T_{\mathbf{k}}^{z}-t_{\bot}^{z}\nonumber\\
&+&[(T_{\mathbf{k}}^{x}-t_{\bot}^{x}-T_{\mathbf{k}}^{z}+t_{\bot}^{z})^2+4(V_{\mathbf{k}}-V_{\mathbf{k}}^{'})^2]^{\frac{1}{2}}\}.\nonumber\\
\end{eqnarray}
Therefore we get two symmetric bands $\alpha$ and $\gamma$, as well as two anti-symmetric ones $\beta$ and $\delta$. The normal Green's function matrix is calculated as $G(k)=(ip_nI-M_\mathbf{k})^{-1}$. Here $I$ is the unit matrix and $k=(\mathbf{k},ip_n)$, with $p_n=(2n-1)\pi T$ and $T$ being the temperature.

The standard multiorbital Hubbard interaction is \cite{scalapino,kubo}
\begin{eqnarray}
\label{honsite}
H_{int}&=&\sum_{i}\sum_{l}\Bigg[U\sum_{o}n_{ilo\uparrow}n_{ilo\downarrow}
+(U^{'}-\frac{J_H}{2})\sum_{o>o^{'}}n_{ilo}n_{ilo^{'}}\nonumber\\
&-&J_H\sum_{o>o^{'}}2\mathbf{S}_{ilo}\cdot\mathbf{S}_{ilo^{'}}
+J^{'}\sum_{o\neq o^{'}}c_{ilo\uparrow}^{\dag}c_{ilo\downarrow}^{\dag}c_{ilo^{'}\downarrow}c_{ilo^{'}\uparrow}\Bigg]\nonumber\\
&=&\sum_{i}\sum_{l}\Bigg[U\sum_{o}n_{ilo\uparrow}n_{ilo\downarrow}
+U^{'}\sum_{o>o^{'}}n_{ilo}n_{ilo^{'}}\nonumber\\
&+&J_H\sum_{o>o^{'}}\sum_{\sigma \sigma^{'}}c_{ilo\sigma}^{\dag}c_{ilo^{'}\sigma^{'}}^{\dag}c_{ilo\sigma^{'}}c_{ilo^{'}\sigma}\nonumber\\
&+&J^{'}\sum_{o\neq o^{'}}c_{ilo\uparrow}^{\dag}c_{ilo\downarrow}^{\dag}c_{ilo^{'}\downarrow}c_{ilo^{'}\uparrow}\Bigg],\nonumber\\
\end{eqnarray}
where $U$, $U^{'}$, $J_H$ and $J^{'}$ are the strength of intra-orbital Coulomb interaction, inter-orbital Coulomb interaction, Hund's coupling and pair hopping, respectively. Here $i$ is the index of the lattice site in one layer, $l$ denotes the layer and $o,o^{'}$ are the orbital indices.

The bare susceptibility $\chi_{0}(q)$ is \cite{anticommutation}
\begin{eqnarray}
\label{X0}
&&\chi_{0}^{s_1s_4,s_2s_3}(q)
=\chi_{0}^{l_1o_1l_4o_4,l_2o_2l_3o_3}(q)\nonumber\\
&=&\frac{1}{2N}\int_{0}^{1/T}d\tau e^{i\omega_n\tau}\sum_{\mathbf{k},\mathbf{k}^{'},\sigma,\sigma^{'}}\nonumber\\
&&\langle T_{\tau}c_{\mathbf{k}+\mathbf{q}l_1o_1\sigma}^{\dag}(\tau)c_{\mathbf{k}l_4o_4\sigma}(\tau) c_{\mathbf{k}^{'}-\mathbf{q}l_2o_2\sigma^{'}}^{\dag}(0)c_{\mathbf{k}^{'}l_3o_3\sigma^{'}}(0)\rangle\nonumber\\
&=&-\frac{T}{N}\sum_k G^{s_2s_4}(k+q)G^{s_1s_3}(k).
\end{eqnarray}
Here $s_i=(l_i,o_i)$ is the combined layer and orbital index with $i=1,\ldots,4$. $N$ is the number of unit cells and $q=(\mathbf{q},i\omega_n)$, with $\omega_n=2n\pi T$. 

Within the random-phase approximation, the spin and charge susceptibilities can be written as \cite{anticommutation}
\begin{eqnarray}
\label{RPA}
\chi_{s}(q)&=&[I-\chi_0(q)U_s]^{-1}\chi_0(q),\nonumber\\
\chi_{c}(q)&=&[I+\chi_0(q)U_c]^{-1}\chi_0(q),
\end{eqnarray}
where the nonzero matrix elements of $U_s$ and $U_c$ are
\begin{eqnarray}
\label{Us}
U_{s}^{lo_1 lo_2,lo_3 lo_4}&=&\begin{cases}
U&\text{$o_1=o_2=o_3=o_4$},\\
U^{'}&\text{$o_1=o_4\neq o_3=o_2$},\\
J_H&\text{$o_1=o_2\neq o_3=o_4$},\\
J^{'}&\text{$o_1=o_3\neq o_2=o_4$},
\end{cases}
\end{eqnarray}
and
\begin{eqnarray}
\label{Uc}
U_{c}^{lo_1 lo_2,lo_3 lo_4}&=&\begin{cases}
U&\text{$o_1=o_2=o_3=o_4$},\\
-U^{'}+2J_H&\text{$o_1=o_4\neq o_3=o_2$},\\
2U^{'}-J_H&\text{$o_1=o_2\neq o_3=o_4$},\\
J^{'}&\text{$o_1=o_3\neq o_2=o_4$},
\end{cases}\nonumber\\
\end{eqnarray}
with $l$ being the layer index and $o_1,o_2,o_3,o_4$ being the orbital indices. At a given $q$, $\chi_0$, $\chi_s$ and $\chi_c$ are all $16\times16$ hermitian matrices. In order to characterize the spin stoner factor, we define $\alpha_s$ as the largest eigenvalue of the matrix $\chi_0(\mathbf{q},i\omega_n=0)U_s$ in the $\mathbf{q}$ space. If $\alpha_s=1$ at a specific $\mathbf{Q}$, then a static long-range magnetic order with a modulation vector $\mathbf{Q}$ will develop. In addition, the momentum structure of the spin fluctuation is manifested as the largest eigenvalue of the matrix $\chi_s(\mathbf{q},i\omega_n=0)$ and is denoted as $\rho_s(\mathbf{q})$. Meanwhile, the largest eigenvalue of the matrix $\chi_0(\mathbf{q},i\omega_n=0)$ is denoted as $\rho_0(\mathbf{q})$, which measures the nesting feature of the band structure.

Close to $T_c$, the linearized Eliashberg equation can be expressed as \cite{anticommutation,eliashberg}
\begin{eqnarray}
\label{Eliashberg}
\lambda \phi^{s_3s_4}(k)&=&-\frac{T}{N}\sum_q\sum_{s_1,s_2,s_5,s_6} G^{s_1s_2}(k-q)G^{s_6s_5}(q-k)\nonumber\\
&&V^{s_6s_4,s_1s_3}(q)\phi^{s_2s_5}(k-q),
\end{eqnarray}
where $\phi(k)$ is the anomalous self energy and the spin singlet pairing interaction is \cite{eliashberg,junhuazhang}
\begin{eqnarray}
\label{V}
V(q)&=&
\frac{1}{2}[3U_s\chi_{s}(q)U_s-U_c\chi_{c}(q)U_c+U_s+U_c].\nonumber\\
\end{eqnarray}
We solve Eq. (\ref{Eliashberg}) by the power method \cite{power method} to find the largest positive eigenvalue $\lambda$ and the corresponding $\phi(k)$ is the preferred pairing function. In this way, the layer-, momentum-, orbital- and frequency-dependence of $\phi(k)$ can all be solved self-consistently. In the iterative process, due to the anti-commutation relation of the fermions, the initial input $\phi(k)$ should satisfy \cite{anticommutation,oddfrequency}
\begin{eqnarray}
\label{anticommutation}
\phi^{s_1s_2}(k)&=&
\phi^{s_2s_1}(-k),
\end{eqnarray}
for spin singlet pairing.
After convergence, the anomalous self energy in the band basis $\Delta(k)$ is calculated as
\begin{eqnarray}
\label{Delta_k}
\Delta(k)&=&Q_{\mathbf{k}}^\dag\phi(k)Q_{-\mathbf{k}}^{*}=Q_{\mathbf{k}}^\dag\phi(k)Q_{\mathbf{k}},
\end{eqnarray}
where $Q_{\mathbf{k}}$ is a unitary matrix that diagonalizes Eq. (\ref{h0}) into Eq. (\ref{eigenvalues}).

Throughout this work, the number of unit cells is set to be $N=64\times64$ and the tight-binding parameters are
$(t_{1}^{x},t_{1}^{z},t_{2}^{x},t_{2}^{z},\epsilon^{x},\epsilon^{z},t_{3}^{xz},t_{4}^{xz},t_{\bot}^{x},t_{\bot}^{z})=$(-0.6003,-0.149,0.0391,-0.0007,1.2193,0.0048,0.2679,-0.072,0.038,-0.999), in units of eV \cite{jiangkun1,jiangkun2,jiangkun3}. The summations over momentum and frequency in Eqs. (\ref{X0}) and (\ref{Eliashberg}) are both done by fast Fourier transformation, where we use $16384$ Matsubara frequencies ($-16383\pi T\leq p_n\leq16383\pi T$ and $-16382\pi T\leq\omega_n\leq16384\pi T$). The temperature is set to be $T=0.007$ eV ($T\approx80$ K). The interaction strength in Eq. (\ref{honsite}) satisfies $U^{'}=U-2J_H$ and we fix $J_H=J^{'}$.

\section{results and discussion}

\begin{figure}
\includegraphics[width=1\linewidth]{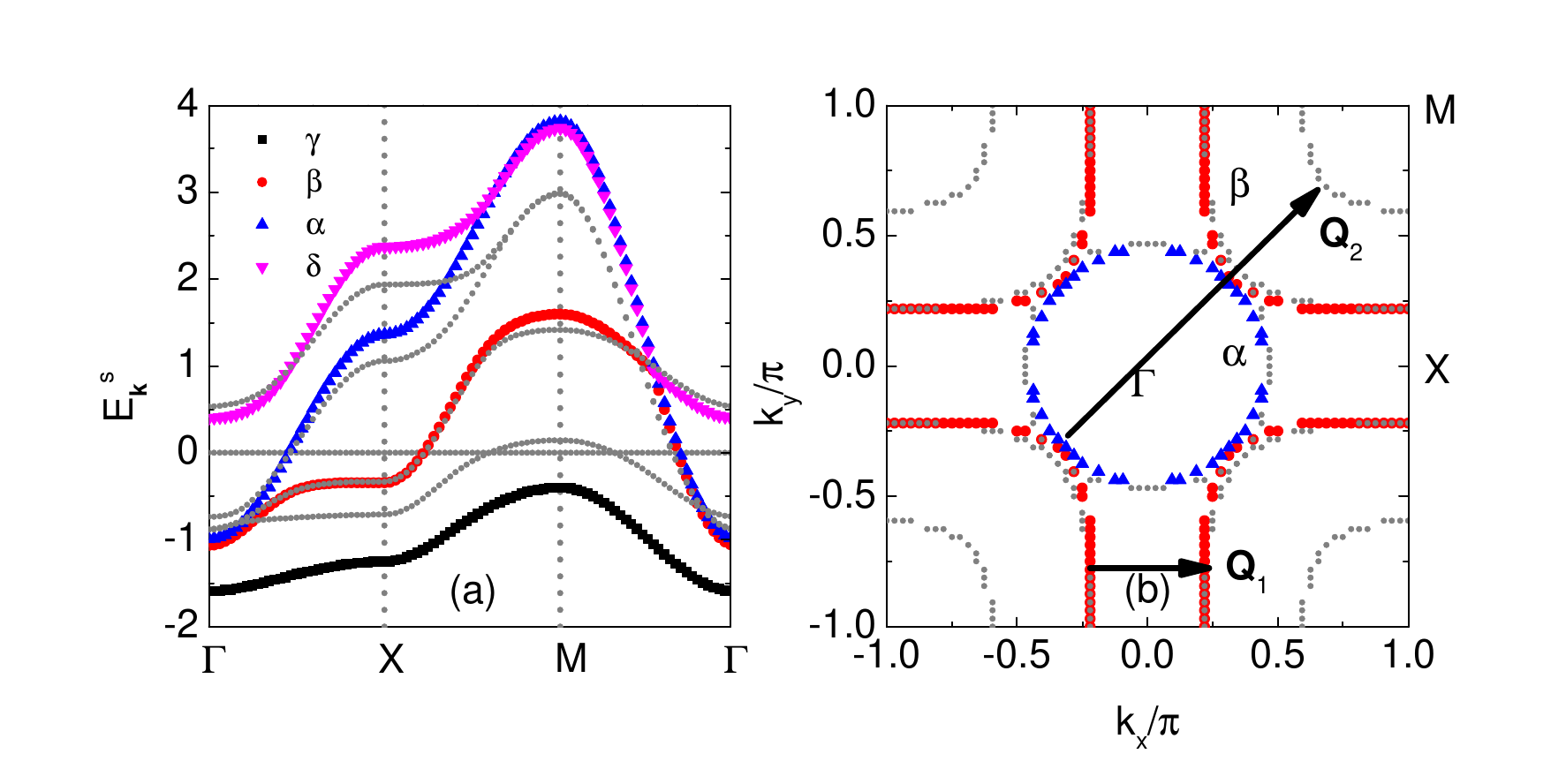}
 \caption{\label{band_structure} (a) The eigenvalues of Eq. (\ref{h0}) along the high-symmetry directions. (b) The corresponding Fermi surface. The gray dotted lines in (a) and (b) are the band structure and Fermi surfaces of the model in Ref. \onlinecite{bilayermodel}.}
\end{figure}

First we show the band structure and Fermi surface of the bilayer two-orbital model in Figs. \ref{band_structure}(a) and \ref{band_structure}(b), respectively. The four bands are labeled as $\alpha,\beta,\gamma,\delta$, among which only two ($\alpha,\beta$) cross the Fermi level and form the Fermi surfaces, while the $\gamma$ band is fully occupied and the $\delta$ band is empty. For comparison, we also plot the band structure and Fermi surfaces of the model in Ref. \onlinecite{bilayermodel} as the gray dotted lines. In Fig. \ref{band_structure}(b), the Fermi surface of each band is approximately determined as $|E_{\mathbf{k}}^{\gamma}|<0.01$ and $|E_{\mathbf{k}}^{\alpha}|,|E_{\mathbf{k}}^{\beta}|<0.03$. We can see, in the present model, the $\gamma$ band shifts downwards to below the Fermi level while its shape stays almost unchanged, the $\beta$ band varies little, the Fermi velocity of the $\alpha$ band increases and so dose the $\delta$ band. In addition, the band width is also enlarged to 5.4, compared to 3.9 of the model in Ref. \onlinecite{bilayermodel}.

\begin{figure}
\includegraphics[width=1\linewidth]{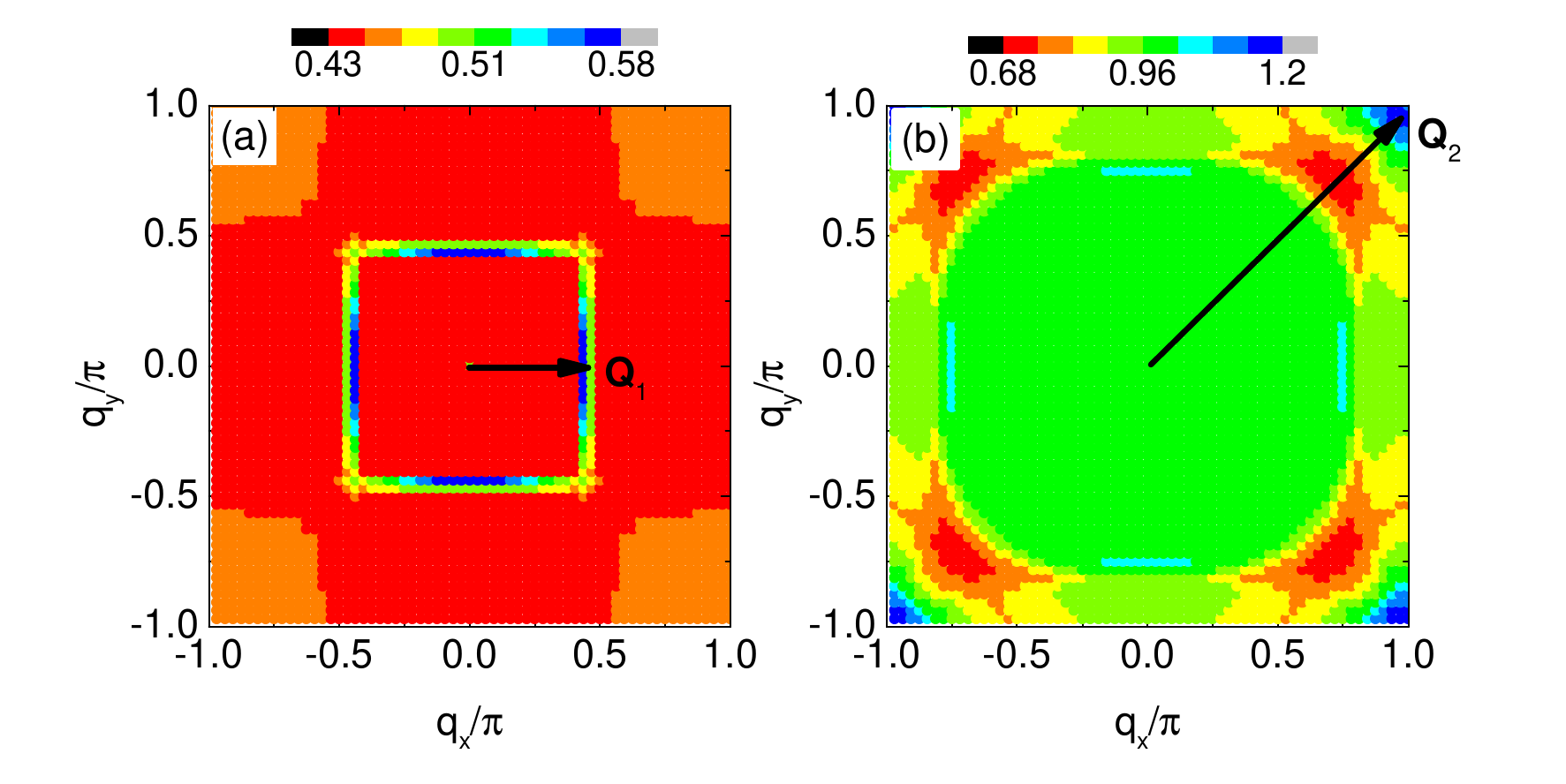}
 \caption{\label{largest_eigenvalue_of_X0} $\rho_0(\mathbf{q})$, defined as the largest eigenvalue of the matrix $\chi_0(\mathbf{q},i\omega_n=0)$, for the present model (a) and the model in Ref. \onlinecite{bilayermodel} (b). }
\end{figure}

Then in Fig. \ref{largest_eigenvalue_of_X0} we show $\rho_0(\mathbf{q})$, defined as the largest eigenvalue of the matrix $\chi_0(\mathbf{q},i\omega_n=0)$. It reflects the nesting feature of the band structure. As we can see from Fig. \ref{largest_eigenvalue_of_X0}(a), in the present model, $\rho_0(\mathbf{q})$ is peaked around $\mathbf{Q}_1=(0,\pm0.5\pi)$ and $(\pm0.5\pi,0)$, originating from the nesting between the straight portions of adjacent $\beta$ Fermi surfaces, connected by the $\mathbf{Q}_1$ wave vector shown in Fig. \ref{band_structure}(b). Besides, around $\mathbf{Q}_2=(\pm\pi,\pm\pi)$, the intensity is medium. This is resulted from the scattering between the top of the $\gamma$ band around $M$ and the $\alpha$ Fermi surface, as well as the bottom of the $\delta$ band around $\Gamma$.
In comparison, for the model in Ref. \onlinecite{bilayermodel}, $\rho_0(\mathbf{q})$ is peaked at $\mathbf{Q}_2$, which is the nesting wave vector between the $\alpha$ and $\gamma$ Fermi surfaces shown in Fig. \ref{band_structure}(b).

\begin{figure*}
\includegraphics[width=1\linewidth]{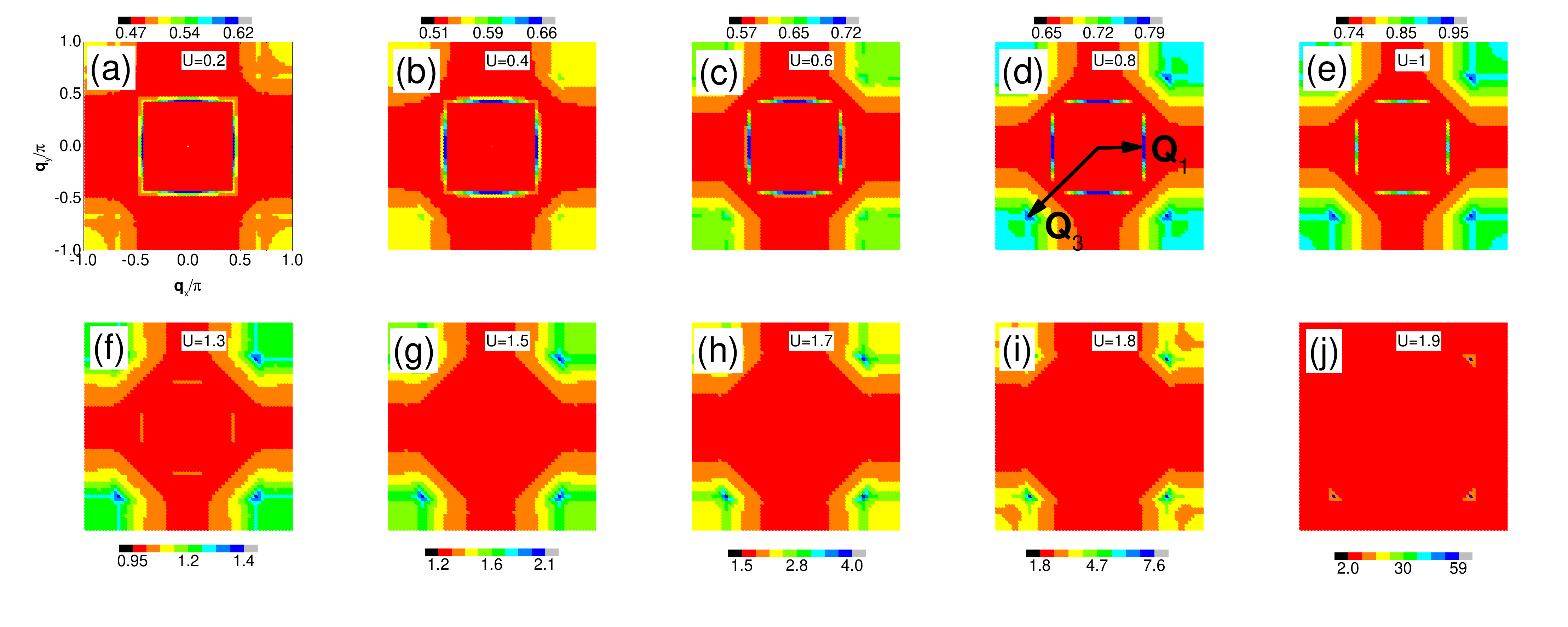}
 \caption{\label{largest_eigenvalue_of_Xs} $\rho_s(\mathbf{q})$, defined as the largest eigenvalue of the matrix $\chi_s(\mathbf{q},i\omega_n=0)$, for different values of $U$ at $J_H=U/6$.}
\end{figure*}

In the presence of the multiorbital Hubbard interaction Eq. (\ref{honsite}), the spin susceptibility $\chi_s$ is enhanced compared to its bare value $\chi_0$. In Fig. \ref{largest_eigenvalue_of_Xs}, we plot $\rho_s(\mathbf{q})$, defined as the largest eigenvalue of the matrix $\chi_s(\mathbf{q},i\omega_n=0)$, for different values of $U$ at $J_H=U/6$. When $U$ is small, $\rho_s(\mathbf{q})$ is similar to $\rho_0(\mathbf{q})$ and is also peaked at $\mathbf{Q}_1$, together with a medium-intensity region around $\mathbf{Q}_2$. The intensity at $\mathbf{Q}_1$ and around $\mathbf{Q}_2$ is enhanced with increasing $U$ [Figs. \ref{largest_eigenvalue_of_Xs}(a) to \ref{largest_eigenvalue_of_Xs}(c)]. At $U=0.8$, the intensity at another wave vector, $\mathbf{Q}_3\approx(\pm0.66\pi,\pm0.66\pi)$, is enhanced significantly and becomes comparable to that of $\mathbf{Q}_1$ [Fig. \ref{largest_eigenvalue_of_Xs}(d)]. As $U$ increases further, the intensity at $\mathbf{Q}_3$ gradually surpasses that of $\mathbf{Q}_1$ and $\rho_s(\mathbf{Q}_3)$ becomes almost divergent at $U=1.9$ [Figs. \ref{largest_eigenvalue_of_Xs}(e) to \ref{largest_eigenvalue_of_Xs}(j)]. Therefore, the momentum structure of the spin fluctuation $\rho_s(\mathbf{q})$ evolves from $\mathbf{Q}_1$-dominated at small $U$ to $\mathbf{Q}_3$-dominated at large $U$.

\begin{figure}
\includegraphics[width=1\linewidth]{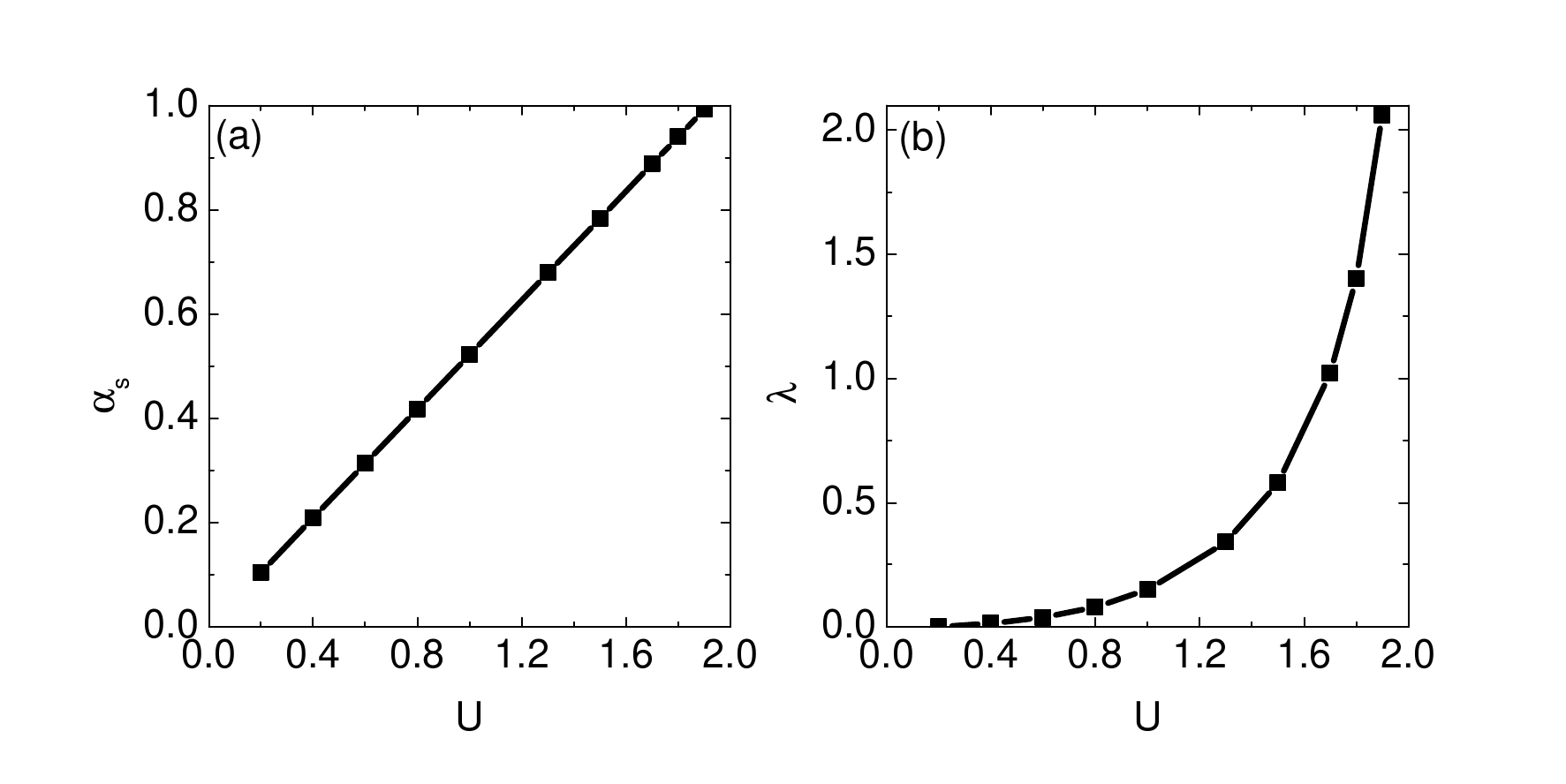}
 \caption{\label{alpha_s&lambda} (a) The spin stoner factor $\alpha_s$ as a function of $U$. (b) The largest positive eigenvalue $\lambda$ of Eq. (\ref{Eliashberg}) as a function of $U$. Here $J_H=U/6$.}
\end{figure}

The spin stoner factor $\alpha_s$, defined as the largest eigenvalue of
the matrix $\chi_0(\mathbf{q},i\omega_n=0)U_s$ in the q space, is shown in Fig. \ref{alpha_s&lambda}(a) for the same values of $U$ and $J_H$ used in Fig. \ref{largest_eigenvalue_of_Xs}. In the present case, the largest eigenvalue of the matrix $\chi_0(\mathbf{q},i\omega_n=0)U_s$ is always located at the $\mathbf{Q}_3$ wave vector, therefore $\alpha_s$ varies linearly with $U$. At $U=1.9$, $\alpha_s\approx1$, thus $\rho_s(\mathbf{Q}_3)$ becomes almost divergent as shown in Fig. \ref{largest_eigenvalue_of_Xs}(j), signifying the tendency of a magnetic order with wave vector $\mathbf{Q}_3$.

\begin{figure*}
\includegraphics[width=1\linewidth]{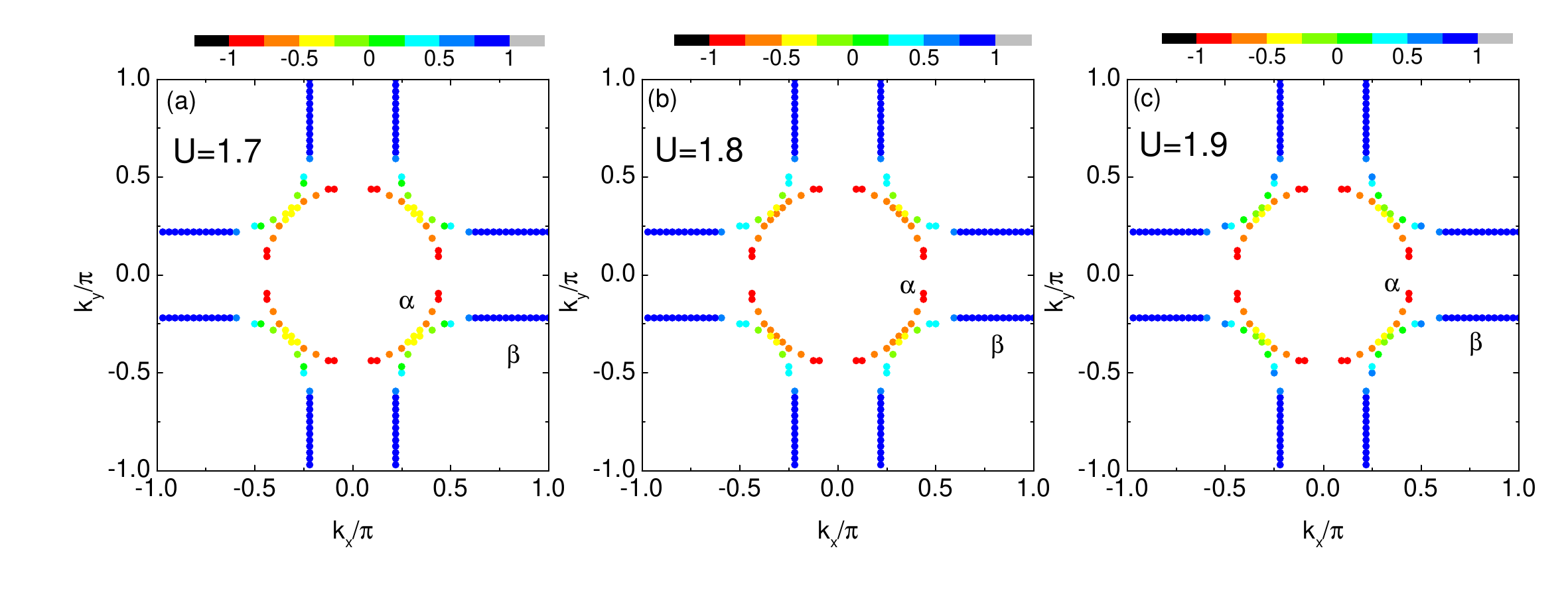}
 \caption{\label{pairing at u=1.7 1.8 1.9} $\Delta^{bb}(\mathbf{k}_F^{b},i\pi T)$ with $b=\alpha,\beta$ at $U=1.7$ (a), $U=1.8$ (b) and $U=1.9$ (c). Here $J_H=U/6$.}
\end{figure*}

\begin{figure}
\includegraphics[width=1\linewidth]{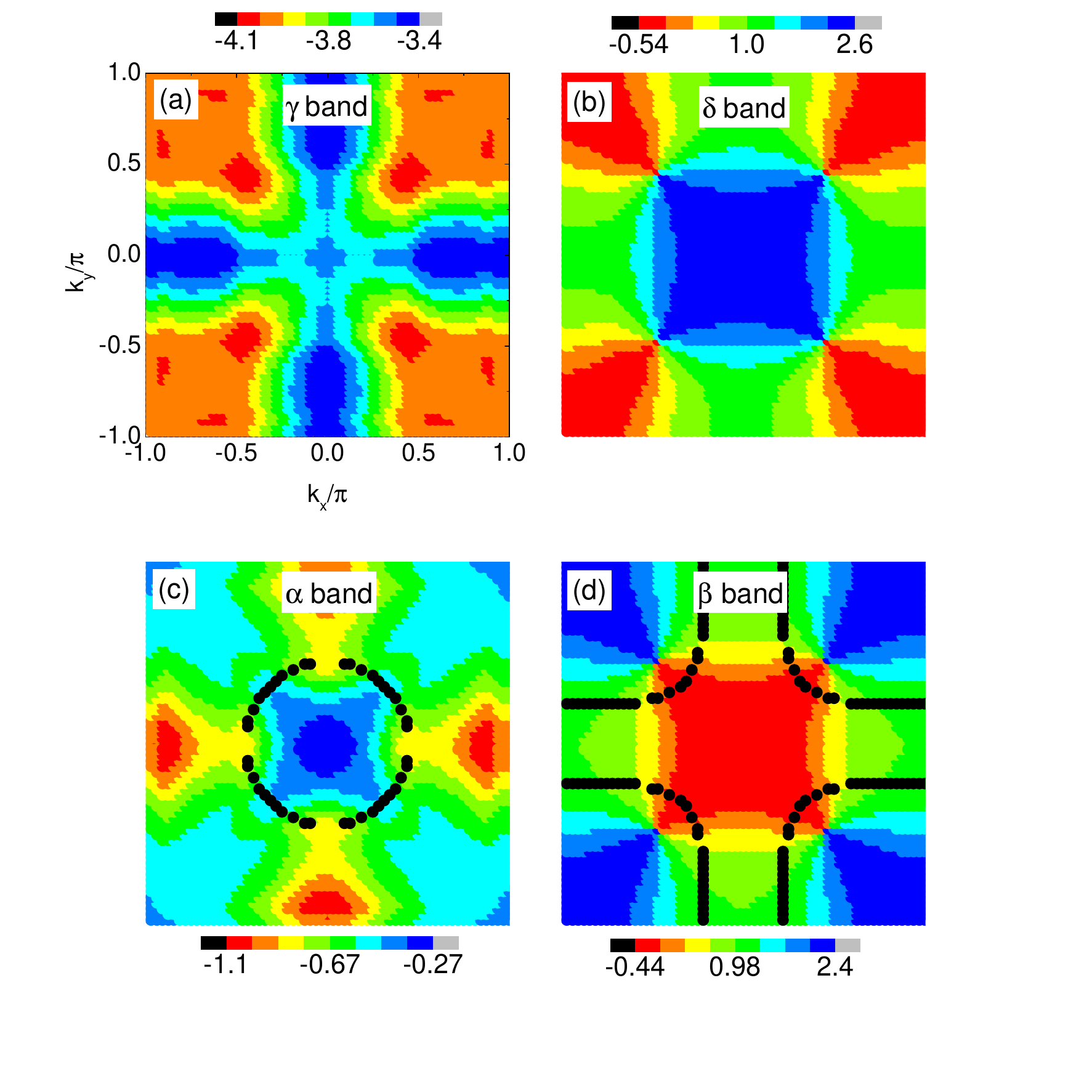}
 \caption{\label{all band u=1.7} $\Delta^{bb}(\mathbf{k},i\pi T)$ with $b=\alpha,\beta,\gamma,\delta$ at $U=1.7$ and $J_H=U/6$. The $\alpha$ and $\beta$ Fermi surfaces are overlaid in (c) and (d), respectively.}
\end{figure}

The largest positive eigenvalue $\lambda$ of Eq. (\ref{Eliashberg}) is plotted in Fig. \ref{alpha_s&lambda}(b), also for the same values of $U$ and $J_H$ used in Fig. \ref{largest_eigenvalue_of_Xs}. At $1.7\leq U\leq1.9$, $\lambda$ exceeds 1 while $\alpha_s$ is less than 1, therefore superconductivity, instead of magnetic order, emerges at this temperature $T=0.007$ eV ($T\approx80$ K). The corresponding eigenvector $\phi(k)$ can then be transformed into the band basis according to Eq. (\ref{Delta_k}). We trace the Fermi momenta $\mathbf{k}_F^{b}$ of each band ($b=\alpha,\beta$ in the present model) and $\Delta^{bb}(\mathbf{k}_F^{b},ip_n)$ at the
lowest positive Matsubara frequency ($p_n = \pi T$) can approximate the pairing function on the Fermi surface. In Fig. \ref{pairing at u=1.7 1.8 1.9} we plot $\Delta^{bb}(\mathbf{k}_F^{b},i\pi T)$ with $b=\alpha,\beta$ at $U=1.7$, 1.8 and 1.9. The pairing functions for the three values of $U$ are similar. $\Delta^{\alpha\alpha}(\mathbf{k}_F^{\alpha},i\pi T)$ is negative while $\Delta^{\beta\beta}(\mathbf{k}_F^{\beta},i\pi T)$ is mostly positive with nodes or gap minima close to the $k_x=\pm k_y$ directions. The pairing symmetry is $s$-wave and since most parts of the $\alpha$ and $\beta$ Fermi surfaces exhibit a sign change, thus it is denoted as $s_\pm$ pairing. We then take the $U=1.7$ case as an example. The pairing function is finite not only on the Fermi surface, but also in the entire Brillouin zone and on all the bands, as shown in Fig. \ref{all band u=1.7}. $\Delta^{\gamma\gamma}(\mathbf{k},i\pi T)$ is negative and has the largest magnitude [Fig. \ref{all band u=1.7}(a)], although this band is below the Fermi level and does not form any Fermi surface. Similarly, $\Delta^{\delta\delta}(\mathbf{k},i\pi T)$ is positive around $\Gamma$ and also has a large magnitude [Fig. \ref{all band u=1.7}(b)], although this band is above the Fermi level and does not form any Fermi surface either. Even on the $\alpha$ and $\beta$ bands that do form the Fermi surfaces, the largest magnitude of the pairing function is not located on their respective Fermi surface, as can be seen from Figs. \ref{all band u=1.7}(c) and \ref{all band u=1.7}(d). In plotting Figs. \ref{pairing at u=1.7 1.8 1.9} and \ref{all band u=1.7}, we normalize $\Delta(\mathbf{k},i\pi T)$ so its largest value on the Fermi surface is 1. We can fit the pairing function to
\begin{eqnarray}
\label{fit}
\phi(\mathbf{k},i\pi T)&=&\begin{pmatrix}
f^{x}_{\mathbf{k}}&f^{'x}_{\mathbf{k}}&f^{xz}_{\mathbf{k}}&f^{'xz}_{\mathbf{k}}\\f^{'x}_{\mathbf{k}}&f^{x}_{\mathbf{k}}&f^{'xz}_{\mathbf{k}}&f^{xz}_{\mathbf{k}}\\
f^{xz}_{\mathbf{k}}&f^{'xz}_{\mathbf{k}}&f^{z}_{\mathbf{k}}&f^{'z}_{\mathbf{k}}\\f^{'xz}_{\mathbf{k}}&f^{xz}_{\mathbf{k}}&f^{'z}_{\mathbf{k}}&f^{z}_{\mathbf{k}}
\end{pmatrix},
\end{eqnarray}
with the nonzero matrix elements being
\begin{eqnarray}
\label{fitcase1}
f^{x}_{\mathbf{k}}&=&-0.32,\nonumber\\
f^{z}_{\mathbf{k}}&=&-0.71,\nonumber\\
f^{'z}_{\mathbf{k}}&=&-2.81.
\end{eqnarray}
Here, $f^{x}_{\mathbf{k}}/f^{z}_{\mathbf{k}}$ is the intra-orbital and intra-layer pairing function in the $x/z$ orbital, while $f^{'x}_{\mathbf{k}}/f^{'z}_{\mathbf{k}}$ is the intra-orbital but inter-layer one. Furthermore, $f^{xz}_{\mathbf{k}}$ is the inter-orbital and intra-layer pairing function and $f^{'xz}_{\mathbf{k}}$ is the inter-orbital and inter-layer one. Equation (\ref{fitcase1}) suggests the pairing is predominantly inter-layer and on the $z$ orbital.

\begin{figure}
\includegraphics[width=1\linewidth]{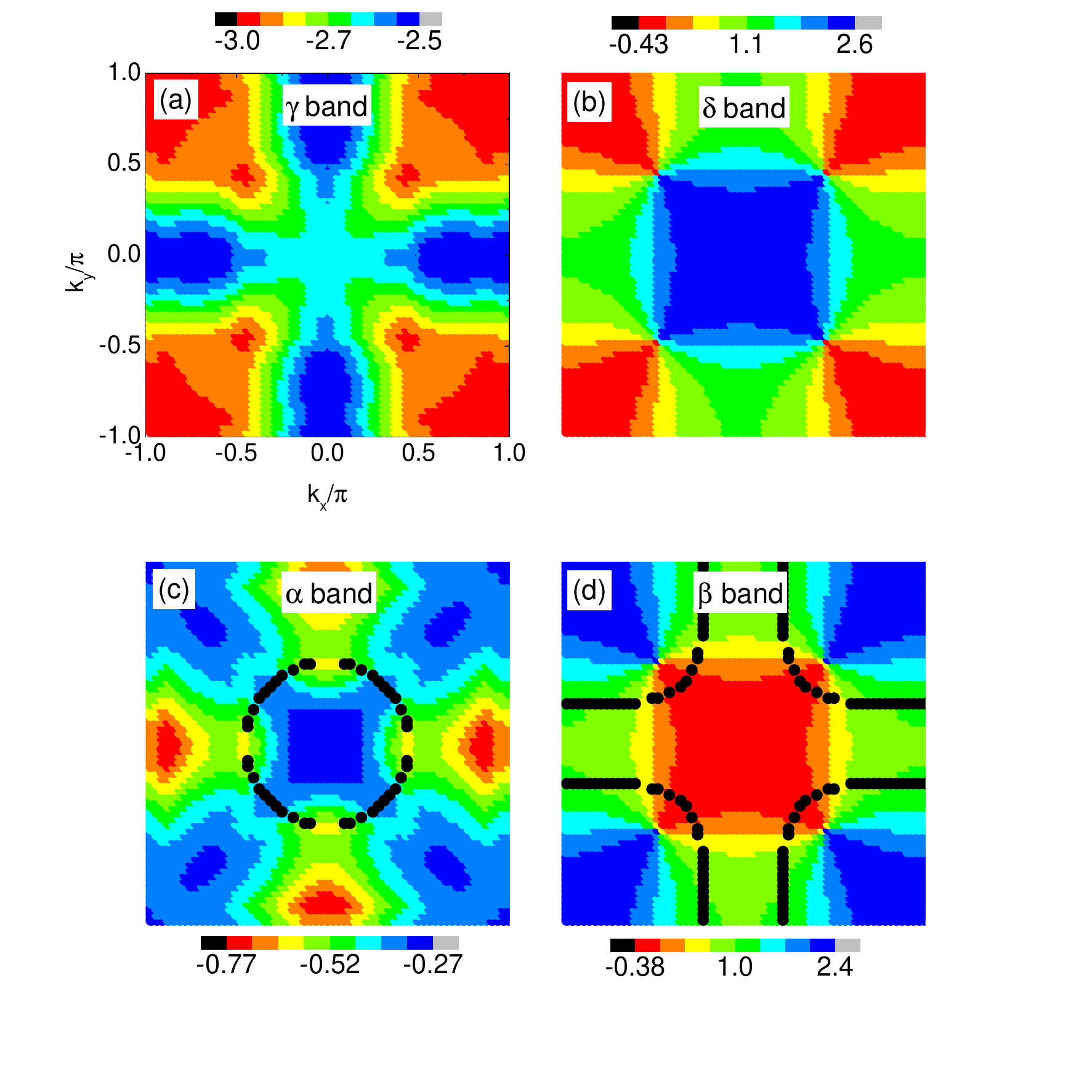}
 \caption{\label{all band u=1.7 neglecting frequency} $\Delta^{bb}(\mathbf{k})$ with $b=\alpha,\beta,\gamma,\delta$ at $U=1.7$ and $J_H=U/6$. The $\alpha$ and $\beta$ Fermi surfaces are overlaid in (c) and (d), respectively.}
\end{figure}

\begin{figure}
\includegraphics[width=1\linewidth]{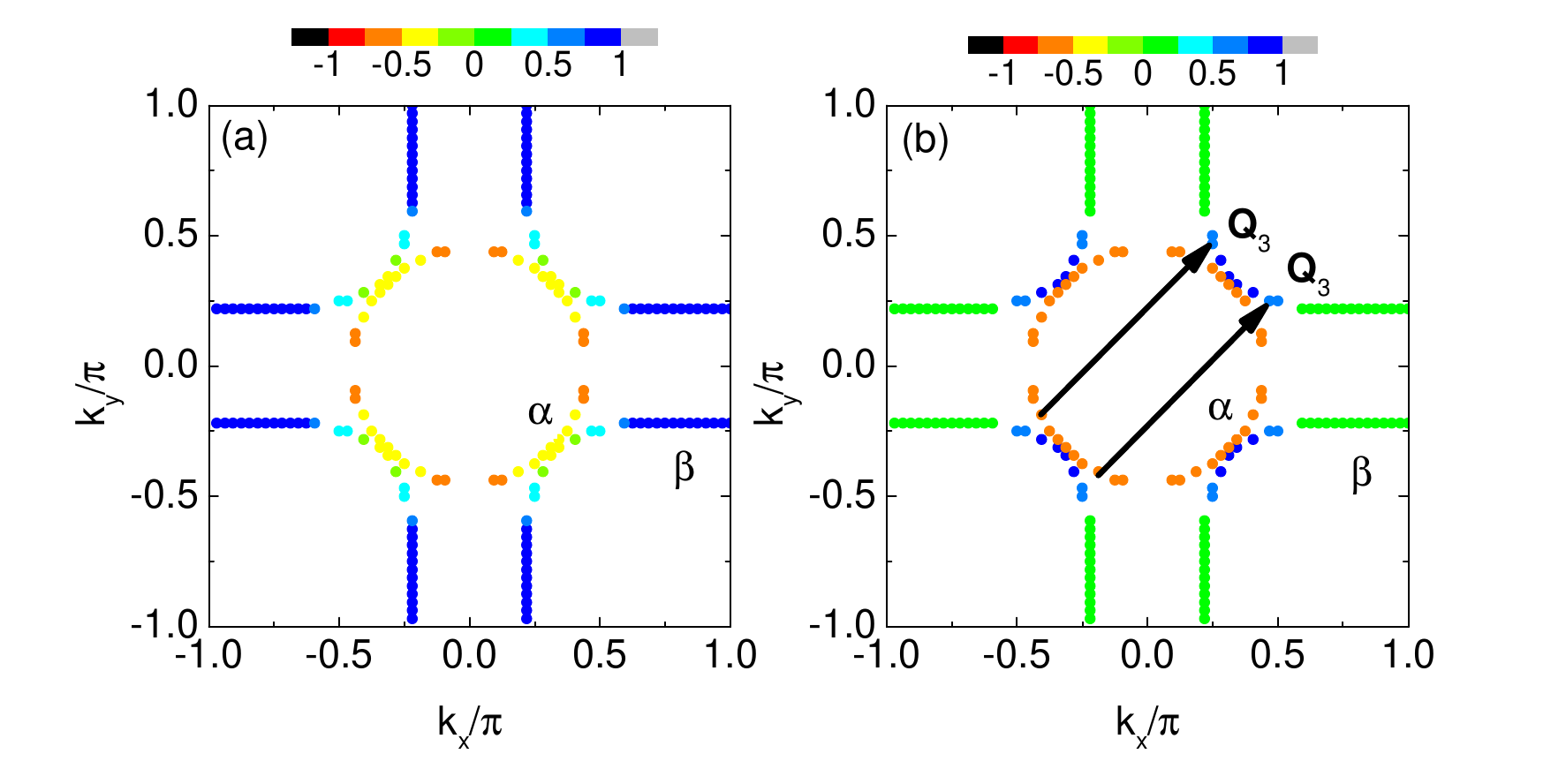}
 \caption{\label{u=1.7 neglecting frequency} $\Delta^{bb}(\mathbf{k}_F^{b})$ with $b=\alpha,\beta$ at $U=1.7$ and $J_H=U/6$. (a) is the result of Eq. (\ref{Eliashberg5}) while (b) is the result of Eq. (\ref{Eliashberg6}).}
\end{figure}

To better understand the pairing function, we analyze the Eliashberg equation further. Eq. (\ref{Eliashberg}) is the standard linearized Eliashberg equation with full orbital-, momentum- and frequency-dependence, which requires a large amount of time and memory to solve in the numerical calculation. Thus some approximations are often made to simplify it. Here we show them in detail. Since $\phi(k)=Q_{\mathbf{k}}\Delta(k)Q_{\mathbf{k}}^{\dag}$ and $Q_{\mathbf{k}}$ is unitary, Eq. (\ref{Eliashberg}) can be written in the band basis as
\begin{eqnarray}
\label{Eliashberg2}
&&\lambda\Delta^{a_1a_2}(\mathbf{k},ip_n)=\frac{T}{N}\sum_{\mathbf{k}^{'},p_{n}^{'}}\sum_{s_1,s_3,s_4,s_6}\sum_{b_1,b_2}\nonumber\\ &&Q_{\mathbf{k}}^{*s_3a_1}Q_{\mathbf{k}}^{s_4a_2}Q_{\mathbf{k}^{'}}^{s_1b_1}Q_{\mathbf{k}^{'}}^{*s_6b_2}V^{s_6s_4,s_1s_3}(\mathbf{k}-\mathbf{k}^{'},ip_n-ip_{n}^{'})\nonumber\\
&&\frac{\Delta^{b_1b_2}(\mathbf{k}^{'},ip_{n}^{'})}{(ip_{n}^{'}-E_{\mathbf{k}^{'}}^{b_1})(ip_{n}^{'}+E_{\mathbf{k}^{'}}^{b_2})}.
\end{eqnarray}
The first approximation is to assume only intra-band pairing, then
\begin{eqnarray}
\label{Eliashberg3}
&&\lambda\Delta^{a_1a_1}(\mathbf{k},ip_n)=\frac{T}{N}\sum_{\mathbf{k}^{'},p_{n}^{'}}\sum_{s_1,s_3,s_4,s_6}\sum_{b_1}\nonumber\\ &&Q_{\mathbf{k}}^{*s_3a_1}Q_{\mathbf{k}}^{s_4a_1}Q_{\mathbf{k}^{'}}^{s_1b_1}Q_{\mathbf{k}^{'}}^{*s_6b_1}V^{s_6s_4,s_1s_3}(\mathbf{k}-\mathbf{k}^{'},ip_n-ip_{n}^{'})\nonumber\\
&&\frac{\Delta^{b_1b_1}(\mathbf{k}^{'},ip_{n}^{'})}{(ip_{n}^{'}-E_{\mathbf{k}^{'}}^{b_1})(ip_{n}^{'}+E_{\mathbf{k}^{'}}^{b_1})}.
\end{eqnarray}
The second approximation is to assume that $\Delta(k)$ is independent of the frequency $p_n$, thus it can be simplified as $\Delta(\mathbf{k})$. The third approximation is to assume that $V(\mathbf{k}-\mathbf{k}^{'},ip_n-ip_{n}^{'})$ can be replaced by $V(\mathbf{k}-\mathbf{k}^{'},i\omega_n=0)$. These two approximations are equivalent to assume that the superconducting pairing is static. In this way, Eq. (\ref{Eliashberg3}) can be simplified as
\begin{eqnarray}
\label{Eliashberg4}
\lambda\Delta^{a_1a_1}(\mathbf{k})&=&\sum_{\mathbf{k}^{'}}\sum_{b_1}A(a_1,\mathbf{k},b_1,\mathbf{k}^{'})\frac{-\tanh(\frac{E_{\mathbf{k}^{'}}^{b_1}}{2T})}{2E_{\mathbf{k}^{'}}^{b_1}}\Delta^{b_1b_1}(\mathbf{k}^{'}),\nonumber\\
\end{eqnarray}
where
\begin{eqnarray}
\label{A}
&&A(a_1,\mathbf{k},b_1,\mathbf{k}^{'})=\frac{1}{N}\sum_{s_1,s_3,s_4,s_6}\nonumber\\
&&Q_{\mathbf{k}}^{*s_3a_1}Q_{\mathbf{k}}^{s_4a_1}Q_{\mathbf{k}^{'}}^{s_1b_1}Q_{\mathbf{k}^{'}}^{*s_6b_1}V^{s_6s_4,s_1s_3}(\mathbf{k}-\mathbf{k}^{'},i\omega_n=0).\nonumber\\
\end{eqnarray}
For singlet pairing, $\Delta^{b_1b_1}(-\mathbf{k}^{'})=\Delta^{b_1b_1}(\mathbf{k}^{'})$, then Eq. (\ref{Eliashberg4}) should be
\begin{eqnarray}
\label{Eliashberg5}
\lambda\Delta^{a_1a_1}(\mathbf{k})&=&\sum_{\mathbf{k}^{'}}\sum_{b_1}L(a_1,\mathbf{k},b_1,\mathbf{k}^{'})\frac{-\tanh(\frac{E_{\mathbf{k}^{'}}^{b_1}}{2T})}{2E_{\mathbf{k}^{'}}^{b_1}}\Delta^{b_1b_1}(\mathbf{k}^{'}),\nonumber\\
\end{eqnarray}
where
\begin{eqnarray}
\label{L}
L(a_1,\mathbf{k},b_1,\mathbf{k}^{'})&=&\frac{1}{2}[A(a_1,\mathbf{k},b_1,\mathbf{k}^{'})+A(a_1,\mathbf{k},b_1,-\mathbf{k}^{'})].\nonumber\\
\end{eqnarray}
In Eq. (\ref{Eliashberg5}), $\mathbf{k}$ and $\mathbf{k}^{'}$ run over the first Brillouin zone, while $a_1$ and $b_1$ run over all the four bands $\alpha,\beta,\gamma,\delta$. It is a simplified eigenvalue equation compared to Eq. (\ref{Eliashberg}) since the frequency dependence is neglected. We solve it by exact diagonalization and plot its eigenvector in Fig. \ref{all band u=1.7 neglecting frequency}. Compared to Fig. \ref{all band u=1.7} we can see, neglecting the frequency dependence results in minor quantitative variation, but not qualitatively. Similarly, on the Fermi surface, the pairing function in Fig. \ref{u=1.7 neglecting frequency}(a) is qualitatively the same as Fig. \ref{pairing at u=1.7 1.8 1.9}(a).

On the other hand, since the function $\frac{\tanh(\frac{E_{\mathbf{k}^{'}}^{b_1}}{2T})}{(2E_{\mathbf{k}^{'}}^{b_1})}$ in Eq. (\ref{Eliashberg5}) is sharply peaked around $E_{\mathbf{k}^{'}}^{b_1}=0$, then the fourth approximation is to assume that Eq. (\ref{Eliashberg5}) can be approximated around the Fermi surface as
\begin{eqnarray}
\label{Eliashberg6}
\lambda\Delta^{a_1a_1}(\mathbf{k}_F^{a_1})&=&\sum_{b_1}\sum_{\mathbf{k}_F^{b_1}}L(a_1,\mathbf{k}_F^{a_1},b_1,\mathbf{k}_F^{b_1})\frac{-\tanh(\frac{E_{\mathbf{k}_F^{b_1}}^{b_1}}{2T})}{2E_{\mathbf{k}_F^{b_1}}^{b_1}}\nonumber\\
&&\Delta^{b_1b_1}(\mathbf{k}_F^{b_1}).\nonumber\\
\end{eqnarray}
Now $a_1$ and $b_1$ run over only the $\alpha$ and $\beta$ bands that form the Fermi surfaces, while $\mathbf{k}_F^{a_1}$ and $\mathbf{k}_F^{b_1}$ are the Fermi momenta of bands $a_1$ and $b_1$, respectively.
Eq. (\ref{Eliashberg6}) is much simpler compared to Eqs. (\ref{Eliashberg}) and (\ref{Eliashberg2}), or even to Eq. (\ref{Eliashberg5}) from the numerical point of view. Therefore a lot of previous studies are based on solving this equation as an eigenvalue problem by considering only a few hundred or thousand Fermi momenta. We also solve it and plot the result in Fig. \ref{u=1.7 neglecting frequency}(b). The pairing symmetry remains $s_\pm$ wave and on those Fermi momenta connected by $\mathbf{Q}_3$, the pairing function changes sign and has the largest magnitude. This pairing function seems very reasonable since it fully utilizes the spin fluctuation structure peaked at $\mathbf{Q}_3$ in Fig. \ref{largest_eigenvalue_of_Xs}(h). However it deviates from those in Figs. \ref{pairing at u=1.7 1.8 1.9}(a) and \ref{u=1.7 neglecting frequency}(a). In Fig. \ref{u=1.7 neglecting frequency}(b), the pairing function on the straight portions of the $\beta$ Fermi surface is now very weak. The deviation originates from those pairing away from the Fermi surfaces, which are neglected in Eq. (\ref{Eliashberg6}). To see this, we derive from Eq. (\ref{Eliashberg5}) that
\begin{eqnarray}
\label{lambda}
\lambda&\propto&\sum_{\mathbf{k},\mathbf{k}^{'}}\sum_{a_1,b_1}L(a_1,\mathbf{k},b_1,\mathbf{k}^{'})\frac{-\tanh(\frac{E_{\mathbf{k}^{'}}^{b_1}}{2T})}{2E_{\mathbf{k}^{'}}^{b_1}}\nonumber\\
&&\Delta^{b_1b_1}(\mathbf{k}^{'})\Delta^{a_1a_1}(\mathbf{k}).
\end{eqnarray}
Here $L(a_1,\mathbf{k},b_1,\mathbf{k}^{'})>0$ comes from the spin-fluctuation-induced singlet pairing interaction, while $-\frac{\tanh(\frac{E_{\mathbf{k}^{'}}^{b_1}}{2T})}{(2E_{\mathbf{k}^{'}}^{b_1})}<0$, therefore, if  $\Delta^{b_1b_1}(\mathbf{k}^{'})\Delta^{a_1a_1}(\mathbf{k})<0$, such terms in the summation are beneficial to a positive $\lambda$. For example, for $b_1=\beta$ and when $\mathbf{k}^{'}$ is located on the straight portions of the $\beta$ Fermi surface, $\frac{\tanh(\frac{E_{\mathbf{k}^{'}}^{\beta}}{2T})}{(2E_{\mathbf{k}^{'}}^{\beta})}\approx\frac{1}{4T}$ is very large and $\Delta^{\beta\beta}(\mathbf{k}^{'})\approx1$, then for $a_1=\gamma$, since $|\Delta^{\gamma\gamma}(\mathbf{k})|$ is large and $\Delta^{\gamma\gamma}(\mathbf{k})<0$, together with $L(\gamma,\mathbf{k},\beta,\mathbf{k}^{'})>0$ for all $\mathbf{k}$, therefore the product of these terms contributes most significantly to $\lambda$. In addition, for $b_1=\delta$ and when $\mathbf{k}^{'}$ is located around $\Gamma$ (near the bottom of the $\delta$ band where $E_{\mathbf{k}^{'}}^{\delta}\approx0.5$), here we have $\frac{\tanh(\frac{E_{\mathbf{k}^{'}}^{\delta}}{2T})}{(2E_{\mathbf{k}^{'}}^{\delta})}\approx1$ and $\Delta^{\delta\delta}(\mathbf{k}^{'})\approx2.6$, then for $a_1=\gamma$ and when $\mathbf{k}$ is located around $M$, $\Delta^{\gamma\gamma}(\mathbf{k})\approx-3$, therefore $\Delta^{\delta\delta}(\mathbf{k}^{'})\Delta^{\gamma\gamma}(\mathbf{k})<0$ and its magnitude is large, together with a large $L(\gamma,\mathbf{k},\delta,\mathbf{k}^{'})$ (since $\mathbf{k}-\mathbf{k}^{'}\approx\mathbf{Q}_2,\mathbf{Q}_3$), such terms contribute second significantly to $\lambda$. That is, when $\mathbf{k}^{'}$ is located around the bottom of the $\delta$ band (the top of the $\gamma$ band), we can find $\mathbf{k}$ near the top of the $\gamma$ band (the bottom of the $\delta$ band), so that $L$ is large, $\frac{\tanh(\frac{E_{\mathbf{k}^{'}}}{2T})}{(2E_{\mathbf{k}^{'}})}\approx1$ and $\Delta(\mathbf{k})\Delta(\mathbf{k}^{'})<0$ with a large magnitude, thus contributing to $\lambda$. However these terms are neglected in Eq. (\ref{Eliashberg6}), leading to the deviation of the pairing function.

\begin{figure}
\includegraphics[width=1\linewidth]{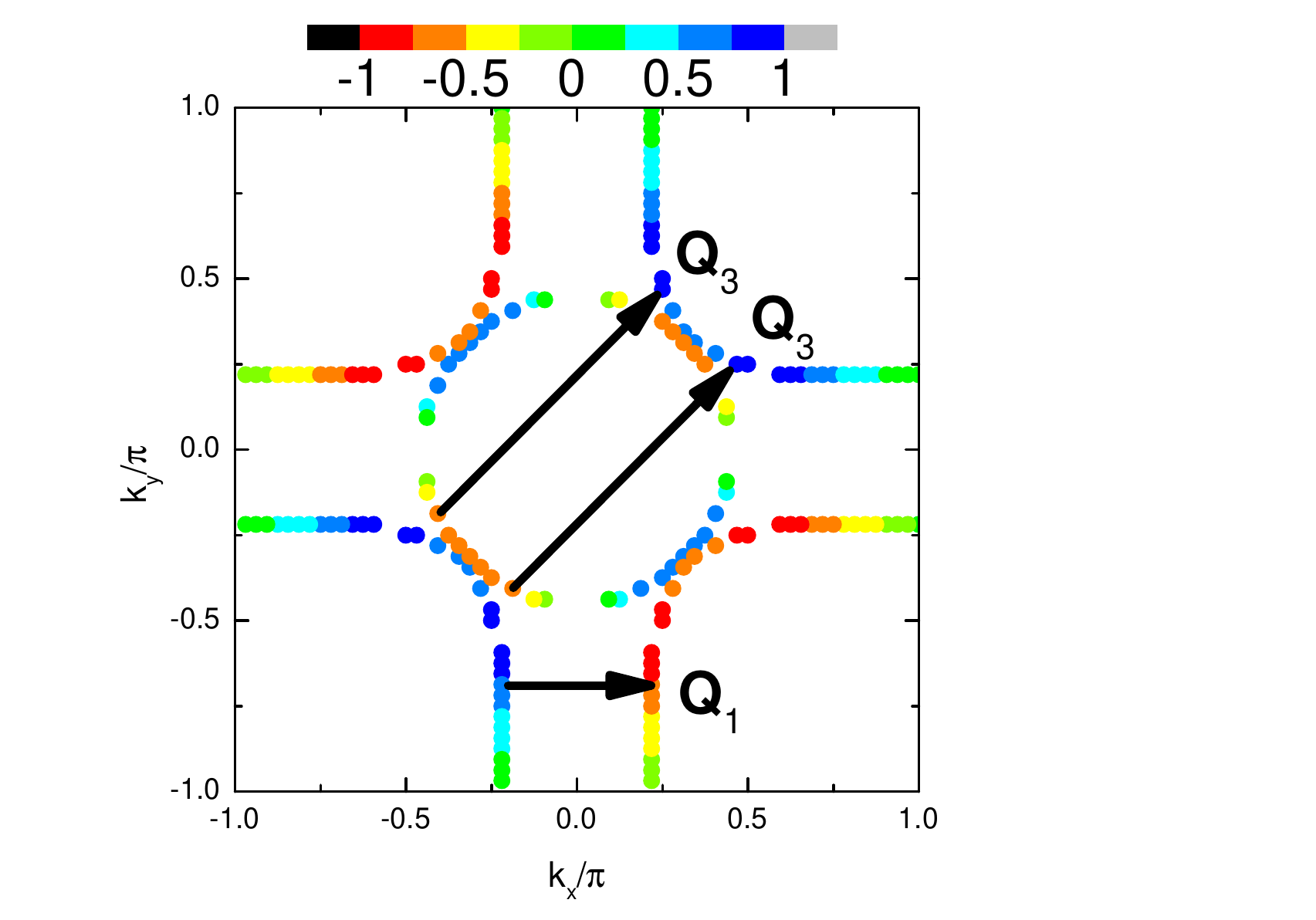}
 \caption{\label{u=0.8 fermi surface pairing} $\Delta^{bb}(\mathbf{k}_F^{b})$ with $b=\alpha,\beta$ at $U=0.8$ and $J_H=U/6$ by solving Eq. (\ref{Eliashberg6}). }
\end{figure}

At other values of $U<1.7$ in Fig. \ref{alpha_s&lambda}(b), $\lambda$ does not reach 1, however the obtained $\phi(k)$ can be viewed as the most probable pairing function if superconductivity indeed emerges. By solving Eq. (\ref{Eliashberg}), we obtained the corresponding $\phi(k)$ for each value of $U$ in Fig. \ref{alpha_s&lambda}(b) and they are all similar to those shown in Figs. \ref{pairing at u=1.7 1.8 1.9} and \ref{all band u=1.7}. Therefore the pairing symmetry and pairing function are both robust to the variation of the interaction strength, although the spin fluctuation structure changes from Figs. \ref{largest_eigenvalue_of_Xs}(a) to \ref{largest_eigenvalue_of_Xs}(j). On the contrary, if we solve Eq. (\ref{Eliashberg6}), then a transition of the pairing symmetry will occur, depending on the value of $U$. When $1.5\leq U\leq1.9$, the pairing function is similar to Fig. \ref{u=1.7 neglecting frequency}(b) and the symmetry is $s_\pm$ wave, since the spin fluctuation is dominated by $\mathbf{Q}_3$, as shown in Figs. \ref{largest_eigenvalue_of_Xs}(g) to \ref{largest_eigenvalue_of_Xs}(j). In contrast, when $0.2<U<1.5$, the pairing symmetry becomes $d_{xy}$-wave shown in Fig. \ref{u=0.8 fermi surface pairing}, where we take $U=0.8$ as an example. Here we can see, on those Fermi momenta connected by $\mathbf{Q}_1$ and $\mathbf{Q}_3$, the pairing function changes sign and is relatively large in magnitude. It is the $\mathbf{Q}_1$ wave vector that changes the pairing symmetry from $s_\pm$ to $d_{xy}$-wave since at this range of $U$, the spin fluctuation has strong intensity at $\mathbf{Q}_1$.

We have also carried out calculations at $J_H=U/4$ and $U/3$. The pairing symmetry remains $s_\pm$ and the pairing function is similar to the $J_H=U/6$ case.

\section{summary}
In summary, we studied the superconducting pairing symmetry based on a newly constructed tight-binding model of La$_3$Ni$_2$O$_7$ under pressure. Compared to the previous bilayer two-orbital model in Ref. \onlinecite{bilayermodel}, the $\gamma$ band sinks below the Fermi level and does not form the Fermi surface anymore. However the pairing symmetry remains $s_\pm$-wave, similar to that obtained by our previous study [Fig. 2(b) in Ref. \onlinecite{gaoyi}]. In the present model, although the $\gamma$ and $\delta$ bands are away from the Fermi level, the superconducting pairing function on them is not zero. Instead, since the top of the $\gamma$ band and bottom of the $\delta$ band are both located at $\sim$500 meV away from the Fermi level, and they are almost nested by the peak structure in the spin fluctuation, thus by forming an anti-phase pairing function on them, these two bands act constructively to increase the value of $\lambda$. The $s_\pm$-wave pairing symmetry is robust against the variation of the interaction strength, although the spin fluctuation structure does not. Finally with detailed derivation and numerical calculation, we demonstrate that the Fermi surface approximated Eliashberg equation may lead to deviation of the pairing symmetry.

\end{document}